\documentstyle[prl,aps,multicols,epsf]{revtex}

\begin{document}

\title{From Strings Theory \\
to the Dark Matter in Galaxies}
\author{Tonatiuh Matos\thanks{%
e-mail:tmatos@fis.cinvestav.mx}}
%EndAName
\address{Departamento de F\'{\i}sica,\\
Centro de Investigaci\'on y de Estudios Avanzados del IPN\\
PO. Box 14--740, 07000 M\'exico D. F., MEXICO}
\maketitle

\begin{abstract}
Starting from the effective action of the low energy limit of Strings
theory, I find an exact solution of the field equations which geodesics
behavie exactly as the trajectories of stars arround of a spiral galaxy.
Here dark matter is of dilatonic origin. It is remarkable that the energy
density of this space-time is the same as the used by astronomers to model
galaxy stability. Some remarks about a universe dominated by dilatons are
pointed out.
\end{abstract}

\draft
\begin{multicols}{2}
\narrowtext

Till some years ago it was believed that in our era, matter is made of
leptons, quarks and gauge bosons. Only theories beyond the Standard Model
predict other exotic particles which can exist at very higher energies,
maybe near of the origin of the universe. But the discovery of the existence
of a great amount of dark matter in galaxies and galaxy clusters could
change our building of how is matter constructed, furthermore it could be
possible that we do not know of what 90\% of the matter of the universe is
made. Let me explain in some lines this affirmation. Since the discovery of
Zwiky and Smith of the necessity of a great amount of wanting matter in the
Coma and Virgo clusters in 1933 \cite{zwicky} in order that these clusters
remain stable, the astronomers have discovered that a great amount of
luminous matter is absent in the galaxies in order to understand their
stability and\ age\ (for a better explanation of the dark matter problem in
galaxies, see the G. Raffelt contribution in this volume). Astronomers have
discovered even a greater amount of wanting luminous matter in most of the
galaxy clusters, since these clusters have also shown to be very stable. In
terms of the reason $\Omega _{x}=\rho _{x}/\rho _{crit}$ between \ a $x$
matter specie and the critical density $\rho _{crit}$ which makes the
universe flat, we can express the contribution of the luminous matter in the
universe by $\Omega _{obs}=(0.003\pm 0.002)/h$ (see for example \cite{schram}%
), which depends on the value of the Hubble constant $h$ in units of 100
km/sec/ Mpc. If we consider the matter needed in the halos of the galaxies
in order to conserve their stability, the mass density of the universe is $%
\Omega _{halos}\sim 0.05$, and considering the matter needed in order to
have stability of the galaxies clusters, the density of the universe grows
to $\Omega _{clost}=0.25\pm 0.10$. These two last densities do not depend on
the value of the Hubble constant. Neutrinos can contribute to the total
density of the universe, nevertheless due to their recently discovered low
mass, their contribution cannot be much grater than the luminous one (see
the contribution of R. Peccei in this volume).

Our actual understanding of the universe is sustained by the Standard Model
of cosmology, namely the Freedman-Robertson-Weaker (FRW) cosmological model.
The predictions of the FRW model is supported for important observations;
the universe expansion, the microwave background and the observations in the
early elements composition in the era of nucleosynthesis. All these three
predictions are supported for a extraordinary coincidences with
observations, therefore it could be very difficult to construct another
cosmological model with so nice features. Remarkable is the fact that the
theoretical predictions of nucleosynthesis do not permit a great amount of
baryonic matter. If the value of the Hubble constant $h=1,$ the permitted
values for the baryonic density implies $0.06<\Omega _{baryon}<0.02.$ If the
Hubble constant $h=0.4$ these values could increase to $0.05<\Omega
_{baryon}<0.12$ (see for example \cite{schram})$.$ In any case this limits
do not permit sufficient baryons for explaining the needed matter in
clusters. {\bf This fact implies that there must exist exotic matter in the
universe}, $i.e.$, there is a great amount of non-baryonic matter in the
cosmos and we do not know its nature. There are many hypothesis about the
nature of this exotic matter. In this lines I want to explain one of this
hypothesis, namely; the scalar field as dark matter in galaxies and in the
universe \cite{fco1}\cite{brena}.

The FRW model contains some problems related with the origin of the
universe, in the quantum mechanic era of the universe. Some of the most
important problems of the FRW model are; the horizon, the flatness problem,
galaxy formation, etc. Some of these problems can be resolved using an
inflationary model of the universe, $i.e.$, introducing a scalar field by
hand into the Einstein field equations. This procedure is preferred by
theoretical physicist because it is elegant and simple. In general the
inflationary model implies that $\Omega =1$, where most of the matter is due
to the scalar field. It is quit remarkable that all the actual most
important unification theories, like the Standard Model of particles, the
Kaluza-Klein and the Strings theories predict the existence of scalar
fields. Scalar fields are needed in order to maintain consistence in the
respective theory. Therefore the question arrays; is it possible that the
wanting matter could have a scalar nature? In this lines I will show that
this seems to be the case, this fact puts the {\bf scalar fields as a good
candidate to be the dark matter in the universe}. I will start supposing
that scalar fields are the dark matter in spiral galaxies. Observational
data show that the galaxies are composed by almost 90\% of dark matter.
Nevertheless the halo contains a larger amount of dark matter, because
otherwise the observed dynamics of particles in the halo is not consistent
with the predictions of Newtonian theory, which explains well the dynamics
of the luminous sector of the galaxy. So we can suppose that luminous matter
does not contribute in a very important way to the total energy density of
the halo of the galaxy at least in the mentioned region, instead the scalar
matter will be the main contributor to it. Luminous matter in galaxies
posses a Newtonian behavior, we expect that only gravitational interactions
are important in them. So, we can perfectly neglect all the other
interactions, I will suppose that only gravitation and scalar interactions
are present. So, the model I am dealing with will be given by the
gravitational interaction modified by a scalar field and a scalar potential.
Then, I start with the effective low energy action of Strings theories with
cosmological constant $\Lambda $ in the Einstein frame \cite{GM}

\begin{equation}
S=\int d^{4}x\sqrt{-g}[-\frac{R}{\kappa _{0}}+2(\nabla \Phi )^{2}+e^{-2\phi
}\Lambda ],  \label{S}
\end{equation}
where $R$ is the scalar curvature, $\Phi $ is the scalar field, $\kappa _{0}=%
\frac{16\pi G}{c^{3}}$ and $\sqrt{-g}$ is the determinant of the metric. I
have carried out a conformal transformation in order to have a more simple
form of the field equations. Action \ (\ref{S}) actually states that an
exponential potential appears in a natural way in this theory.

On the other hand, the exact symmetry of the halo is stills unknown, but it
is reasonable to suppose that the halo is symmetric with respect to the
rotation axis of the galaxy. Here I let the symmetry of the halo as general
as I can, so I choose it to be axial symmetric. Furthermore, the rotation of
the galaxy do not affect the motion of test particles around the galaxy,
dragging effects in the halo of the galaxy should be too small to affect the
tests particles (stars) traveling around the galaxy. Hence, in the region of
interest we can suppose the space-time to be static, given that the circular
velocity of stars (like the sun) of about 230 Km/s seems not to be affected
by the rotation of the galaxy and we can consider a time reversal symmetry
of the space-time. The most general static and axial symmetric metric
compatible with this action, written in the Papapetrou form is 
\begin{equation}
ds^{2}=\frac{1}{f}[e^{2k}(dzd\bar{z})+W^{2}d\phi ^{2}]-f\ c^{2}dt^{2},
\label{2}
\end{equation}
where $z:=\rho +i\zeta $ and $\bar{z}:=\rho -i\zeta $ and the functions $f,\
W$ and $k$ depend only on $\rho $ and $\zeta $. This metric represents the
symmetries posted above.

An exact solution of the field equations derived from the action (\ref{S})
in Boyer-Lindquist coordinates $\rho =\sqrt{r^{2}+b^{2}}\sin \theta $, $%
\zeta =r\cos \theta $ reads \cite{matos1}\cite{matos2}

\begin{eqnarray}
ds^{2} &=&\frac{1+\frac{b^{2}\cos ^{2}\theta }{r^{2}}}{f_{0}r_{0}}(\frac{%
dr^{2}}{1+\frac{b^{2}}{r^{2}}}+r^{2}\ d\theta ^{2})+  \nonumber \\
&&\frac{r^{2}+b^{2}\sin ^{2}\theta }{f_{0}r_{0}}d\phi ^{2}-f_{0}c^{2}\frac{%
r^{2}+b^{2}\sin ^{2}\theta }{r_{0}}dt^{2}  \label{DM}
\end{eqnarray}
The effective energy density $\mu _{DM}$ of (\ref{DM}) is given by the
expression

\begin{equation}
\mu _{DM}=\frac{1}{2}V(\Phi )=\frac{2f_{0}r_{0}}{\kappa _{0}(r^{2}+b^{2}\sin
^{2}\theta )}  \label{muDM}
\end{equation}
The energy density (\ref{muDM}) coincides with that required for a galaxy to
explain the rotation curves of test particles in its halo, but in our model,
this energy density is produced by the scalar field and the scalar field
potential, that is, this dark matter is produced by a $\Phi $ particle.

In what follows I study the circular trajectories of a test particle on the
equatorial plane taking the space-time (\ref{2}) as the background. The
motion equation of a test particle in the space-time (\ref{2}) can be
derived from the Lagrangian 
\begin{equation}
{\cal L}=\frac{1}{f}[e^{2k}(\left( \frac{d\rho }{d\tau }\right) ^{2}+\left( 
\frac{d\zeta }{d\tau }\right) ^{2})+W^{2}\left( \frac{d\phi }{d\tau }\right)
^{2}]-f\ c^{2}\left( \frac{d\ t}{d\tau }\right) ^{2}.  \label{lgeo}
\end{equation}
This Lagrangian contains two constants of motion, the angular momentum per
unit of mass $B$ and the total energy of the test particle $A.$ In terms of
the metric components and the test particle velocity $v=(\dot{\rho},\dot{%
\varsigma},\dot{\phi})$ I obtain $A^{2}=c^{4}f^{2}/(f-\frac{v^{2}}{c^{2}}).$
For a circular trajectory at the equatorial plane $\dot{\varsigma}=$ $\dot{%
\rho}=0$ the equation of motion is $B^{2}\ f/W^{2}-A^{2}/c^{2}f=-c^{2}.$
This last equation determines the circular trajectories of the stars of the
galaxy. Using these equations I obtain an expression for $B$ in terms of $%
v^{2}$, $B^{2}=v^{2}/(f-\frac{v^{2}}{c^{2}})W^{2}/f\sim v^{2}W^{2}/f^{2},$
since $v^{2}<<c^{2}$. From this equation one concludes that for our solution
(\ref{DM}) $v^{2}=f_{0}^{2}B^{2}$, $i.e.$ 
\begin{equation}
v_{DM}=f_{0}B,  \label{Ma}
\end{equation}
where I call $v\rightarrow v_{DM}$ the circular velocity due to the dark
matter alone.

Let us model the circular velocity profile of a spiral galaxy by the
function 
\begin{equation}
v_{L}^{2}=v^{2}(R_{opt})\beta {\frac{1.97\,x^{1.22}}{({x}^{2}+{0.78}%
^{2})^{1.43}}}  \label{vL}
\end{equation}
which is the approximate model for the Universal Rotation Curves proposed by
Persic $et.al.$ \cite{persic} where $\beta =v_{L}(R_{opt})/v(R_{opt.}).$ One
obtains a typical profile of the circular velocity due to the luminous
matter of a spiral galaxy \cite{begeman}. With this velocity it is now easy
to calculate the angular momentum (per unity of mass) of the test particle $%
B=v_{L}D,$ where $D$ is the distance between the center of the galaxy and
the test particle. For our metric, $D=\int ds$ $=\sqrt{%
(r^{2}+b^{2})/f_{0}r_{0}}$. Finally, using equation (\ref{Ma}) I find the
profile of the dark matter velocity. The results are shown in fig. 1 for
some galaxies. We see that the correspondence with typical circular
velocities profiles given in the literature \cite{begeman}\cite{vera} is
excellent.

\begin{figure*}
\label{fig1}
\leftline{ \epsfxsize=40mm \epsfbox{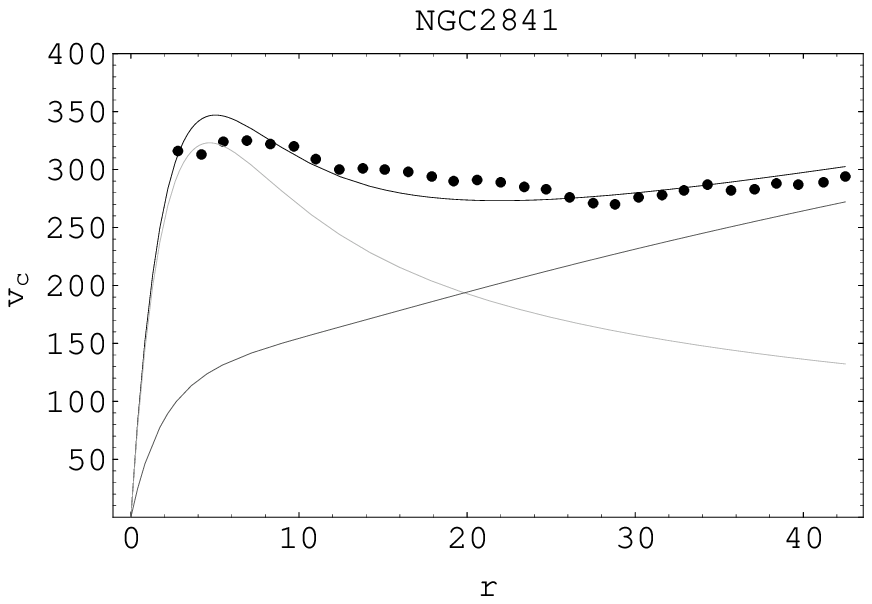} \epsfxsize=40mm
\epsfbox{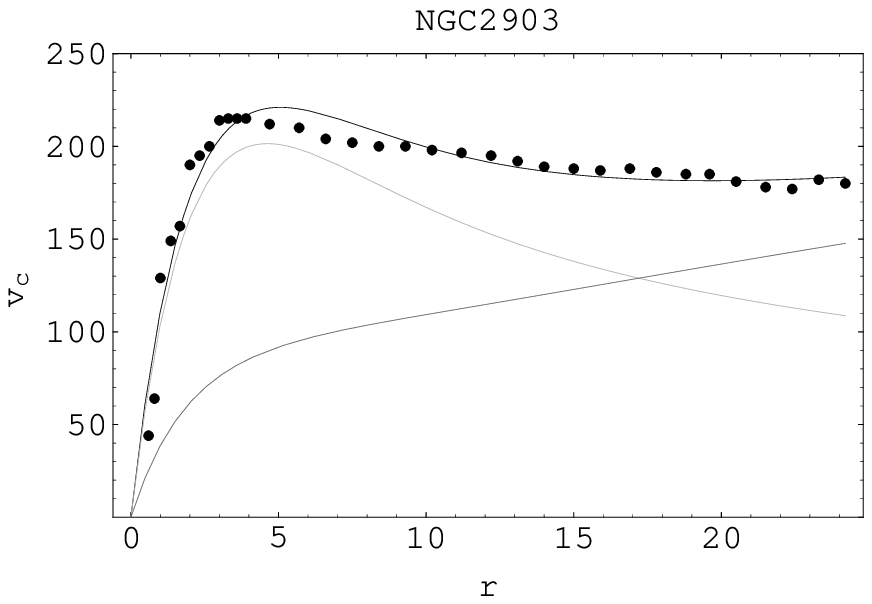}}
\leftline{ \epsfxsize=40mm \epsfbox{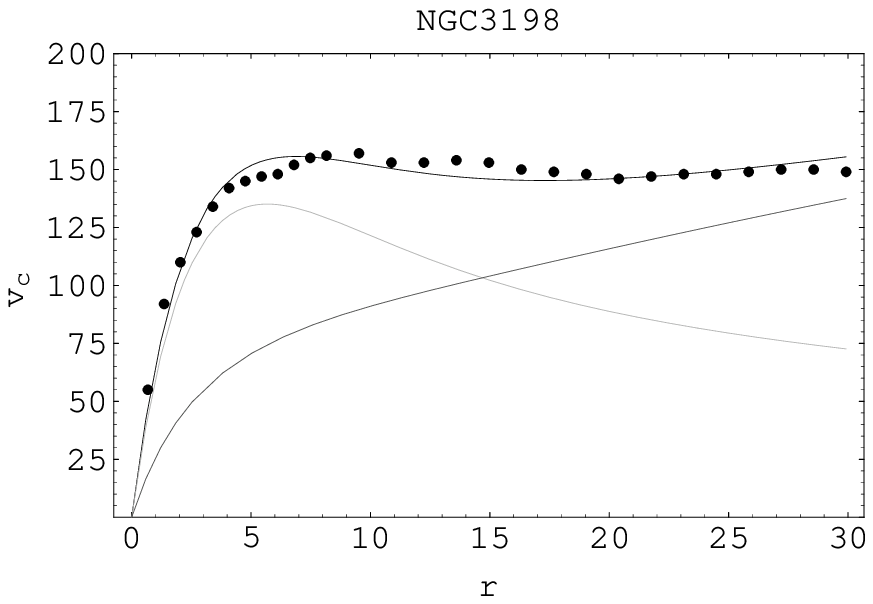} \epsfxsize=40mm
\epsfbox{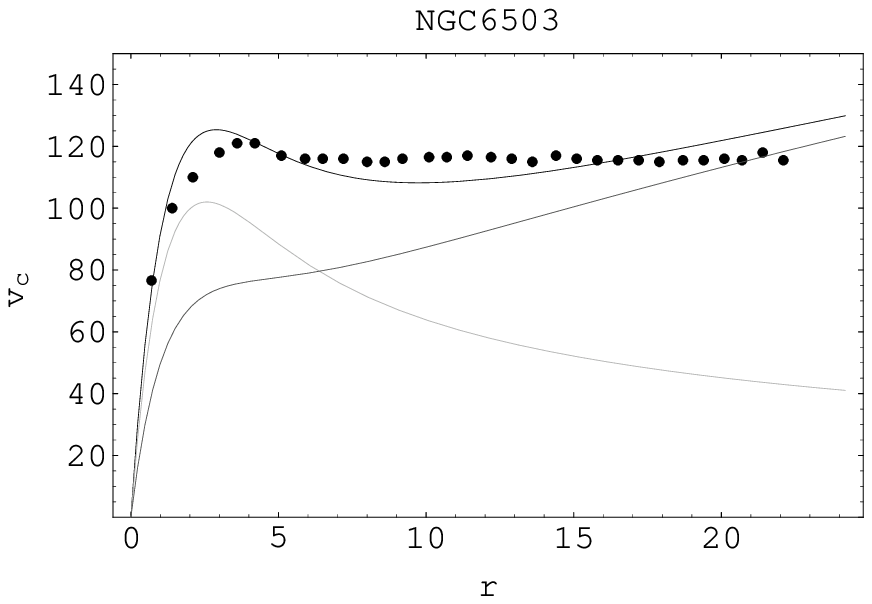}}
\caption{The circular velocity profiles of four galaxies. In this
Black lines represents the total circular velocity ($v$),
midle-grays is the contribution of the dark matter to the total velocity
($v_{DM}$) and the light-grays curves is the contribution of luminous
matter ($v_L$); finally the dots represent the observational data. The
units are in (Km/s) in the vertical axis and in (Kpc) in the
horizontal one.}
\end{figure*}

The crucial point for having the circular velocity $v_{DM}=f_{0}B$ is that $%
f\sim W$ in the solution (\ref{DM}). But this fact remains unaltered after
conformal transformations in the metric ${\hat{d}s}^{2}=F(\Phi )ds^{2}$, so
that the circular velocity $v_{DM}$ remains the same for all theories and
frames related with metric (\ref{2}) by conformal transformations. This
point is very important. In order to derive action (\ref{S}) from the
effective action of the low energy limit of Strings theory, I have carried
out a conformal transformation, so this result is valid also for this last
action. But then the result should be valid for any theory conformaly
equivalent to action (\ref{S}).

This result has some very interesting consequences \cite{fco}. If this
result is true, Scalar fields not only exist, but they represent 90\% of the
matter in the universe. This result and the inflationary models tell us that
scalar fields are the most important part of matter in nature, they
determine the structure of the universe. After the big bang, they inflated
the universe; soon after they gave mass to the particles; later they
concentrate maybe because of scalar field condensation, provoking that
baryonic matter density fluctuations and forming stars, galaxies and galaxy
clusters. Scalar fields can clarify why galaxies formed so soon after the
recombination era, they condensate during the radiation era forming the
arena that formed the galaxies. The question why nature use only the spin 1
and spin 2 fundamental interactions over the simplest spin 0 interactions
becomes clear here. This result tell us that in fact nature have preferred
the spin 0 interaction over the other two ones, scalar field interactions
determine the cosmos structure. This result give also a limit for the
validity of the Einstein's equations, they are valid at local level;
planets, stars, star-systems, but they are not more valid at galactic or
cosmological level, a scalar field interaction must be added to the original
equations. This result could be the first contact of higher dimensional
theories like the Kaluza-Klein or the Strings one with reality, furthermore,
it could be the first trace to demonstrate the existence of extra dimensions
in nature.

This work was partially supported by CONACyT M\'{e}xico, grant 3697-E.%
\newline

\end{multicols}
\end{document}